\documentclass[11pt]{article}
\usepackage{moriond,epsfig}
\usepackage[a4paper,left=0.9in,right=0.9in,top=0.90in,bottom=0.90in]{geometry}
\usepackage{amsmath}
\usepackage{amssymb}
\usepackage{xspace}

\newcommand{\jet}{\ensuremath{\,\mathrm{jet}}\xspace}

\newcommand{\nbar}{\ensuremath{\bar{n}}}
\newcommand{\nLO}{\nbar \text{LO}\xspace}
\newcommand{\nNLO}{\nbar \text{NLO}\xspace}
\newcommand{\NNLO}{\text{NNLO}\xspace}
\newcommand{\LS}{\mathrm{LS}}
\newcommand{\GeV}{\ensuremath{\,\mathrm{GeV}}\xspace}
\newcommand{\TeV}{\ensuremath{\,\mathrm{TeV}}\xspace}
\newcommand{\miss}{\ensuremath{\,\mathrm{miss}}\xspace}

\def\be{\begin{equation}}
\def\ee{\end{equation}}
\def\bea{\begin{eqnarray}}
\def\eea{\end{eqnarray}}

\def\bit{\begin{itemize}}
\def\eit{\end{itemize}}

\begin{document}
\vspace*{2.5cm}
\title{%
WZ and W+jets production at large transverse momenta beyond NLO
}

\author{SEBASTIAN SAPETA}

\address{%
  Institute for Particle Physics Phenomenology, Durham University,\\
  South Rd, Durham DH1 3LE, UK
}

\maketitle
\vspace{-18em}
\begin{flushright}
  IPPP/13/33 \\
  DCPT/13/66
\end{flushright}
\vspace{15em}
\abstracts{
  We present a study of higher order QCD corrections beyond NLO to processes
  with electroweak vector bosons. We focus on the regions of high transverse
  momenta of commonly used differential distributions. We employ the LoopSim
  method, combined with NLO packages, VBFNLO and MCFM, to merge the NLO samples
  with different multiplicities, in order to compute the dominant part of the
  NNLO corrections at high $p_T$.
  We find that these corrections are indeed substantial, in the 30\%-100\%
  range, for a number of experimentally relevant observables. For other
  observables, they lead to significant reduction of scale uncertainties.
}

\section{Introduction}

The production of electroweak vector bosons forms one of the most important
class of Standard Model (SM) processes. 
W boson in association with jets is a background to single and pair top
production, diboson production, Higgs production as well as to searches
for physics beyond the standard model (BSM). The same is true for the
processes with two electroweak bosons in the final state, like WZ production,
which, in addition, are sensitive to anomalous triple gauge boson coupling
(TGC).
The above processes are also interesting in their own right as they
provide important tests of quantum chromodynamics (QCD).
 
\begin{figure}[t]
\begin{center}
  \begin{minipage}[b]{0.26\linewidth}
    \begin{center}
    \psfig{figure=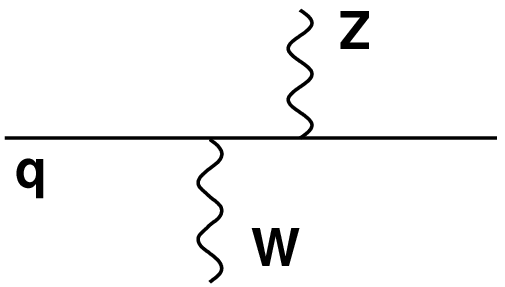, height=1.8cm}

    (LO) \\
    \end{center}
  \end{minipage}
  \hspace{20pt}
  \begin{minipage}[b]{0.26\linewidth}
    \begin{center}
    \psfig{figure= 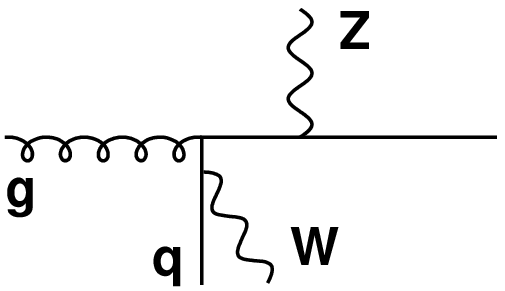,    height=1.8cm}

    (NLO)
    \end{center}
  \end{minipage}
  \hspace{20pt}
  \begin{minipage}[b]{0.26\linewidth}
    \begin{center}
    \psfig{figure= 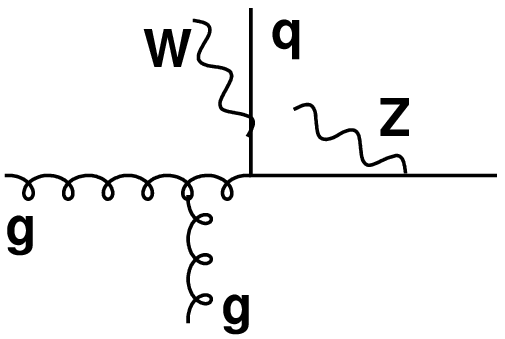,    height=1.8cm}

    (NNLO)
    \end{center}
  \end{minipage}
\end{center}
\caption{
  Example diagrams contributing to WZ production at LO, NLO and NNLO.
}
\label{fig:diagrams}
\end{figure}

In this proceedings, we present a study of WZ and W+jets processes, at the LHC
energies, at approximate next-to-next-to-leading order (NNLO) in QCD. The
motivation to go beyond the next-to-leading order (NLO) for those processes
comes from the fact that the NLO corrections for WZ and W+jets turn out to be
sizable for a number of important distributions at high transverse momentum. 
This corrections come about due to new production channels and new topologies
absent at leading order (LO) and appearing only at NLO. An example, for the case
of WZ, is shown in Fig.~\ref{fig:diagrams}. At leading order, the production of
dibosons is possible only via $q\bar q$ channel. At NLO, the new $qg$ channel,
with enhanced partonic luminosity, opens up and dominates the LO contribution.
Similarly, at LO, only back-to-back WZ configurations are possible whereas at
NLO, an electroweak boson can recoil against a parton and the other boson can be
soft or collinear, which brings logarithmic enhancements for a number of
distributions.

Because the NLO corrections often turn out to dominate the leading order, it is
of great importance to try to assess the NNLO corrections, to check the
convergence of the perturbative series, and to obtain precise and stable
results.  As shown in the right diagram of Fig.~\ref{fig:diagrams}, in the case
of WZ production, one can also expect genuinely new sub-processes and topologies
appearing for the first time at NNLO.

\section{Details of the calculations}

To compute the dominant part of the NNLO QCD corrections to WZ and W+jets
processes, we used the LoopSim method~\cite{Rubin:2010xp} together with the NLO
packages VBFNLO~\cite{Arnold:2012xn,Campanario:2010hp} and MCFM.~\cite{mcfm} 

LoopSim allows for a consistent merging of WZ and WZj NLO
samples~\cite{Campanario:2010hp} to obtain the result for WZ at \nNLO, where
$\bar n$ denotes the approximate 2-loop contribution determined from the real
and 1-loop parts of WZj at NLO.  Similarly, Wj and Wjj NLO samples are used to
compute the results for W+jets at \nNLO. The method is based on unitarity and it
is supposed to work best for processes and observables with large NLO K-factors
from new topologies, as $\sigma_{\nNLO} - \sigma_{\NNLO} \sim \frac{1}{K}$.
LoopSim has one parameter, $R_\LS$, which affects the way the procedure assigns
branching structure to the original NLO sample. By default, we set $R_\LS=1$ and
vary it by $\pm0.5$, to probe the related uncertainties. As we shall see, they
are much smaller then the factorization and renormalization scale uncertainties
at high $p_T$.

In the entire study, regardless of the order, we used the MSTW NNLO
2008~\cite{Martin:2009iq} PDFs. 

\section{Results}

The WZ production process was studied~\cite{Campanario:2012fk} with the
following cuts: The charged leptons were required to be hard and central:
$p_{T,\ell}\ge 15(20)$, for $\ell$ coming from Z(W), and $|y_{\ell}|\le 2.5$.
The missing transverse energy had to satisfy the cut $E_{T,\text{miss}} > 30
\GeV$.  The same-flavor lepton pair mass had to lie in the window $60 <
m_{l^+l^-} < 120 \GeV$. 
The final state partons with $|y_p|\le 5$ were clustered to jets with the
anti-$k_t$ algorithm~\cite{Cacciari:2008gp}, as implemented in
FastJet~\cite{Cacciari:2005hq,fastjet:2005}, with the radius $R=0.45$. For the
central value of the factorization and renormalization scales, we chose
$\mu_{F,R}= \frac12 \sum p_{T,\text{partons}} + \frac12 \sqrt{p_{T,W}^2+m_W^2}+
\frac12 \sqrt{p_{T,Z}^2+m_Z^2}$. All WZ results correspond to the sum of
contributions from two unlike flavor decay channels, $ee\mu\nu_\mu$ and $\mu\mu
e\nu_e$, and both W$^+$Z and W$^-$Z production channels.

\begin{figure}[t]
\begin{center}
\psfig{figure= 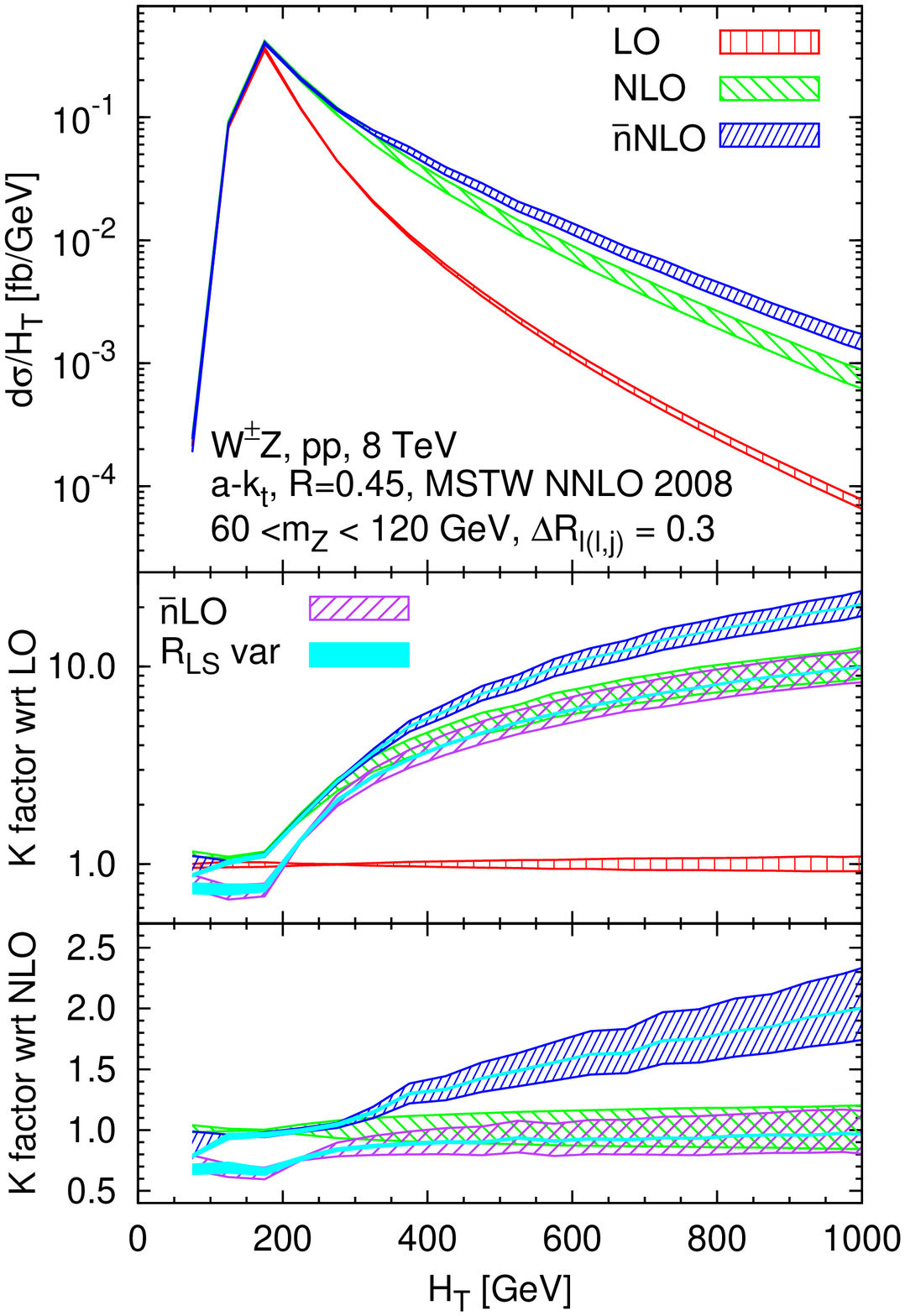,   height=8.7cm}
\hspace{50pt}
\psfig{figure= 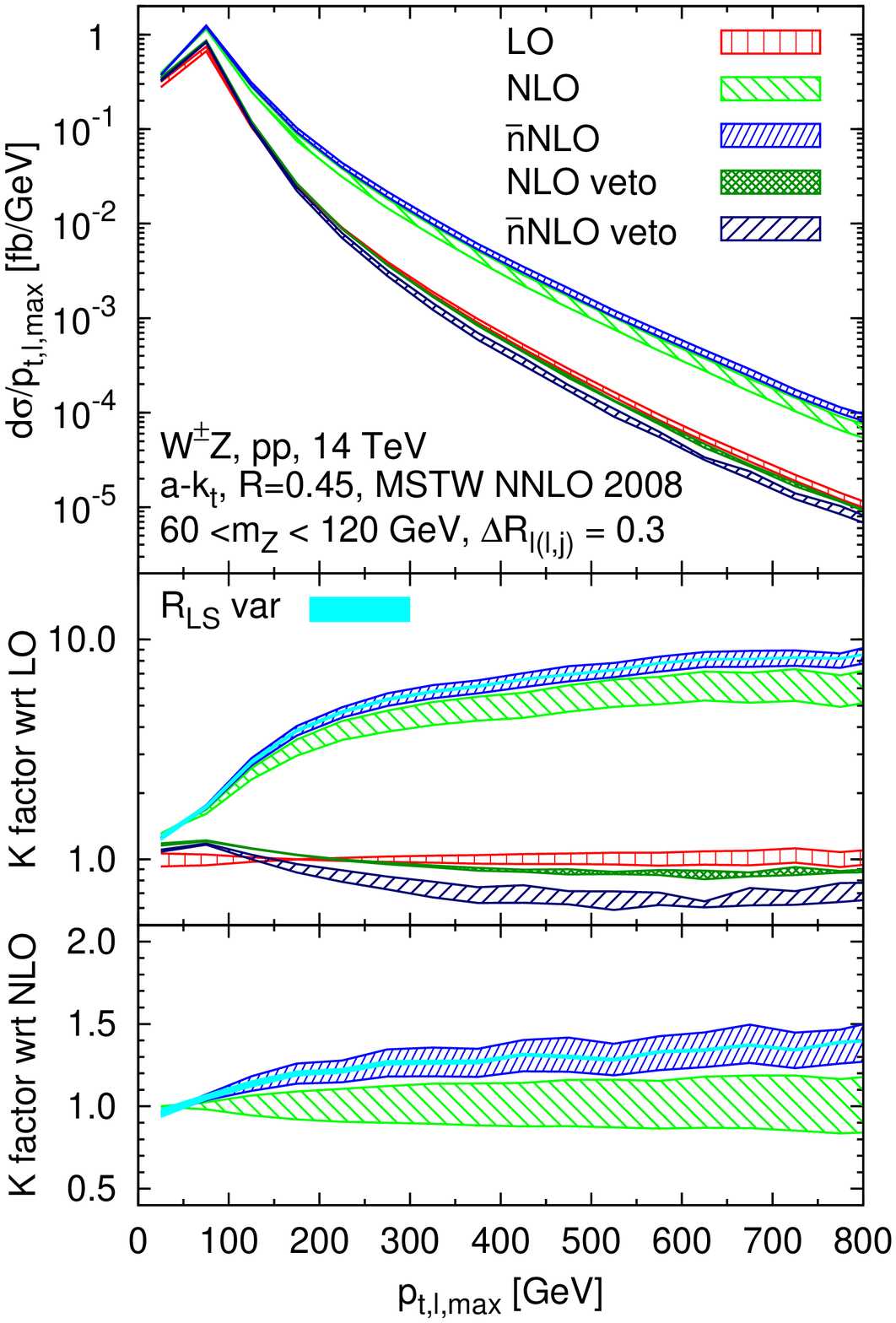, height=8.7cm}
\end{center}
\caption{
  WZ production at the LHC.  Differential cross sections and K factors for
  $H_T$, defined in Eq.~(\ref{eq:HT}), at $\sqrt{s}=8\, \text{TeV}$ (left) and
  for the $p_T$ of the hardest lepton at $\sqrt{s}=14\, \text{TeV}$ (right).
  The bands correspond to varying $\mu_F=\mu_R$ by factors 1/2 and 2 around the
  central value. The cyan solid bands give the
  uncertainty related to $R_\text{LS}$ varied between 0.5 and 1.5.
  The distribution is a sum of contributions from two unlike flavor decay
  channels.
}
\label{fig:wz}
\end{figure}

Fig.~\ref{fig:wz} (left) shows the differential distribution, at $\sqrt{s}=8\,
\text{TeV}$, of the effective mass observable defined as
\begin{equation}
H_{T} = \sum  p_{T,\text{jets}} + \sum  p_{T,l} +
E_{T,\text{miss}}\,.
\label{eq:HT}
\end{equation}
As a first check, we computed the $H_T$ distribution at \nLO, which can be
directly compared with the exact NLO result. As we see in Fig.~\ref{fig:wz},
\nLO matches very well the NLO at high $H_T$, providing the correct prediction
for the large K factor, of the order of 10. The $H_T$ observable is therefore
very well suited to be studied with LoopSim. The \nNLO result shown in
Fig.~\ref{fig:wz} yields up to 100\% correction with respect to NLO. We see that
the $R_\LS$ uncertainty is negligible at high $H_T$ and the scale uncertainty
decreases only a little at \nNLO. The latter is related to the fact that this
observable favours new topologies that enter only at NNLO and are computed
with LO accuracy (cf. the right diagram of Fig.~\ref{fig:diagrams}).

In Fig.~\ref{fig:wz} (right), we present the distribution of the lepton with
maximum $p_T$ at $\sqrt{s}=14\, \text{TeV}$. Also here, the \nNLO corrections
are large and beyond the NLO scale uncertainty already at 200 GeV. In addition,
for this observable, we show the results with veto on the jets with $p_T > 50
\GeV$.
We observe that the \nNLO corrections with veto are negative, go beyond the NLO
scale uncertainty, and exhibit larger uncertainties due to renormalization and
factorization scale variation than the NLO result. The latter indicates that the
small scale uncertainty at NLO was partially accidental. Overall, these results
clearly illustrate that the veto procedure should be used with great care as it
has a non-trivial effect on the convergence of the perturbative series.


The W+jets process was studied with the cuts that match the ATLAS
measurement.~\cite{Aad:2012en}
The charged leptons were required to have $p_{T,\ell}\ge 20 \GeV$ and
$|y_{\ell}|\le 2.5$. The missing transverse energy had to be above
$E_{T,\text{miss}} > 25 \GeV$. The transverse mass~\cite{Aad:2012en} of the W
was required to be greater than $40 \GeV$. Only events with
anti-$k_t$~\cite{Cacciari:2008gp}, R=0.4 jets with $p_{T,\jet} > 30 \GeV$ and
$|y_{\jet}| < 4.4$ were accepted. Finally, for each jet, its distance to the
lepton $\Delta R(\ell,\jet)$ had to be greater than 0.5, otherwise this jet was
removed from the event.
For the central value of the factorization and the renormalization scale, we
chose $\mu_{F,R} = \frac12 \hat H_T = \frac12 \left(\sum
p_{T,\text{partons}}+\sum p_{T,\text{leptons}} \right)$.

\begin{figure}[t]
\begin{center}
\psfig{figure= 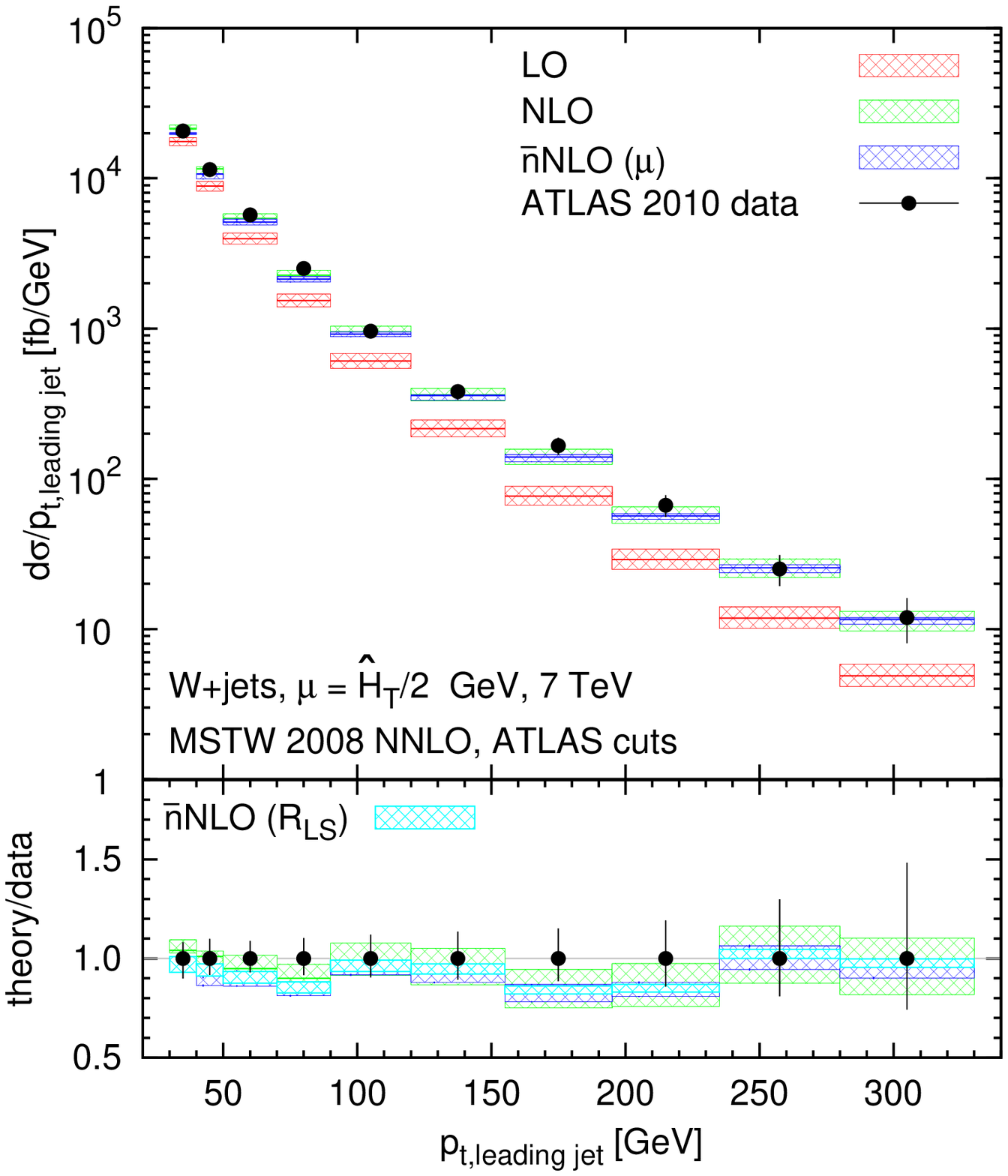, height=7.5cm}
\hspace{30pt}
\psfig{figure= 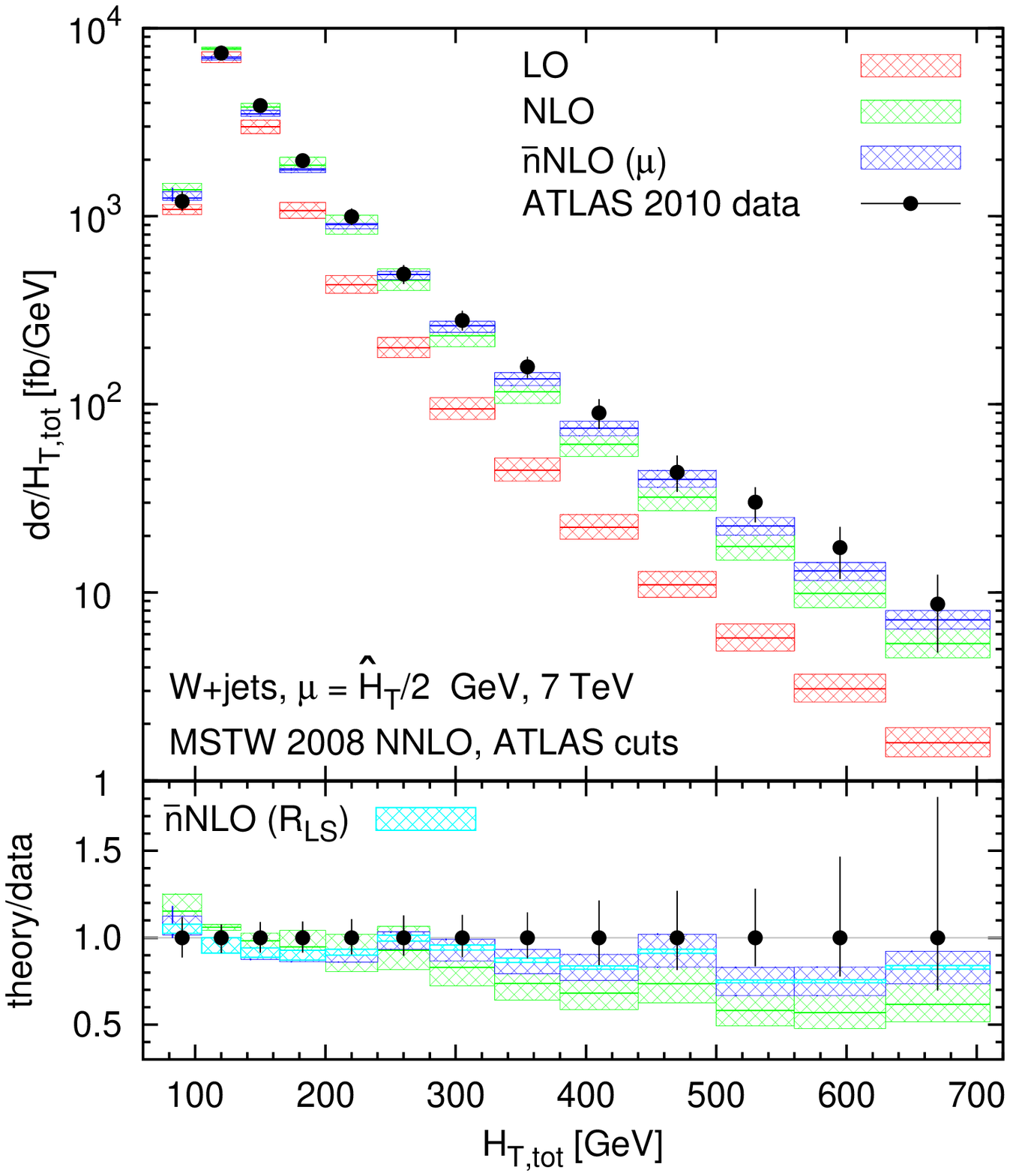,  height=7.5cm}
\end{center}
\caption{
  W+jets production at the LHC, $\sqrt{s} = 7 \TeV$.
  Differential cross sections as functions of the $p_T$ of the hardest jet
  (left) and $H_T$ (right) at LO, NLO and
  \nNLO.
  The theoretical results are corrected for non-perturbative effects and are
  compared to the ATLAS data.
  The bands correspond to varying $\mu_F=\mu_R$ by factors 1/2 and 2 around the
  central value. The cyan solid bands give the
  uncertainty related to $R_\LS$ varied between 0.5 and 1.5.  
}
\label{fig:wjets}
\end{figure}

In Fig.~\ref{fig:wjets} we show the differential distributions of the
transverse momentum of the hardest jet~(left) and the scalar sum of the traverse
momenta of jets, leptons and missing energy, $H_{T}$~(right), defined previously
in Eq.~(\ref{eq:HT}). 
The results correspond $\sqrt{s} = 7 \TeV$ and are sums of
contributions from $W^+$ and $W^-$ for a single lepton decay channel $W\to \ell
\nu$. The theoretical predictions, computed at LO, NLO and \nNLO,
were corrected for non-perturbative effects.~\cite{Aad:2012en}
 
In the case of the $p_T$ of the leading jet, we see a substantial reduction of
scale uncertainty at \nNLO, while the result stays within the NLO band. 
Hence, that observable comes under control at \nNLO as no new channel or
topologies appear at this order. We also note that the $R_\LS$ uncertainty is
always smaller that the scale uncertainty and it decreases with increasing
$p_T$.
For the $H_{T}$ distribution, the \nNLO result goes beyond the NLO uncertainty
band for $H_{T}> 300 \GeV$ and the corrections are up to 30\% with respect to
NLO. The $R_\LS$ uncertainty is negligible above 300 GeV.  
The large \nNLO correction to $H_{T}$ is a result of the third jet coming from  
the initial state radiation. This jet adds a small contribution to
$H_{T}$ but, because the spectrum is steeply falling, the enhancement in
the distribution is substantial. Altogether, the \nNLO result, by including
configurations with three partons in the final state, describes the $H_{T}$
data significantly better than NLO.

\section{Conclusions}

We used LoopSim together with VBFNLO and MCFM to compute approximate NNLO
corrections to processes with vector bosons: WZ and W+jets production. Our
results, referred to as \nNLO, are expected to account for the dominant part of
the NNLO QCD corrections in some high $p_T$ distributions.

In the case of WZ production, we found sizable effects due to \nNLO for a range
of experimentally relevant observables: $H_T$, $p_{T,\ell, \max}$,
$E_{T,\miss}$.~\cite{Campanario:2012fk} They all show non-trivial kinematic
dependencies and go beyond NLO uncertainty bands. For W+jets, the $H_T$-type
observables exhibit large \nNLO corrections. On the other hand, $p_T$ of the
leading jet, which gets large correction at NLO, nicely converges at \nNLO and
shows significant reduction of scale uncertainty.

In conclusion, the QCD corrections beyond NLO to process with electroweak
bosons are in many cases sizable and should be taken into account in precision
studies as well as in searches for new physics.

\section*{Acknowledgments}
We thank the organizers of the 48$^{\rm th}$ Rencontres de Moriond for the
opportunity to present these results and for financial support.
The studies described here were performed in collaboration with Francisco
Campanario, Daniel Maitre and Gavin Salam.


\end{document}